\documentclass[journal=jacsat,manuscript=article]{achemso}
\usepackage{chemformula} % Formula subscripts using \ch{}
\usepackage[T1]{fontenc} % Use modern font encodings
\usepackage{braket}

\newcommand*{\Eres}{250 meV}

\author{Jin Bakalis}
\author{Sergii Chernov}
\affiliation{Department of Chemistry, Stony Brook University, Stony Brook, NY 11794, USA.}

\author{Ziling Li}
\affiliation{Department of Physics, The Ohio State University, Columbus, OH 43210, USA.}

\author{Alice Kunin}
\affiliation{Department of Chemistry, Stony Brook University, Stony Brook, NY 11794, USA.}
\altaffiliation{Current address: Department of Chemistry, Princeton University, Princeton, NJ 08544, USA.}

\author{Zachary H. Withers}
\affiliation{Department of Physics and Astronomy, Stony Brook University, Stony Brook, NY 11794, USA.}

\author{Shuyu Cheng}
\affiliation{Department of Physics, The Ohio State University, Columbus, OH 43210, USA.}

\author{Alexander Adler}
\affiliation{Department of Physics and Astronomy, Stony Brook University, Stony Brook, NY 11794, USA.}

\author{Peng Zhao}
\altaffiliation{Current address: Photonics Industries International Inc, Ronkonkoma, NewYork, 11779}
\author{Christopher Corder}
\affiliation{Department of Chemistry, Stony Brook University, Stony Brook, NY 11794, USA.}

\author{Michael G. White}
\affiliation{Department of Chemistry, Stony Brook University, Stony Brook, NY 11794, USA.}
\alsoaffiliation{Chemistry Division, Brookhaven National Laboratory, Upton, NY 11973 USA.}

\author{Gerd Sch\"onhense}
\affiliation{Johannes Gutenberg-Universit\"at, Institut f\"ur Physik, D-55099 Mainz, Germany.}

\author{Xu Du}
\affiliation{Department of Physics and Astronomy, Stony Brook University, Stony Brook, NY 11794, USA.}

\author{Roland K. Kawakami}
\affiliation{Department of Physics, The Ohio State University, Columbus, OH 43210, USA.}

\author{Thomas K. Allison}
\affiliation{Department of Chemistry, Stony Brook University, Stony Brook, NY 11794, USA.}
\alsoaffiliation{Department of Physics and Astronomy, Stony Brook University, Stony Brook, NY 11794, USA.}

\email{thomas.allison@stonybrook.edu}

\title[An \textsf{achemso} demo]
  {Momentum-space Observation of Optically Excited Non-Thermal Electrons in Graphene with Persistent Pseudospin Polarization.}

\abbreviations{IR,NMR,UV}

\begin{document}

% \begin{tocentry}

% Some journals require a graphical entry for the Table of Contents.
% This should be laid out ``print ready'' so that the sizing of the

% Inside the \texttt{tocentry} environment, the font used is Helvetica
% 8\,pt, as required by \emph{Journal of the American Chemical

% The surrounding frame is 9\,cm by 3.5\,cm, which is the maximum
% permitted for  \emph{Journal of the American Chemical Society}
% graphical table of content entries. The box will not resize if the
% content is too big: instead it will overflow the edge of the box.

% This box and the associated title will always be printed on a
% separate page at the end of the document.

% \end{tocentry}

\begin{abstract}
  The unique optical properties of graphene, with broadband absorption and ultrafast response, make it a critical component of optoelectronic and spintronic devices. 
  Using time-resolved momentum microscopy with high data rate and high dynamic range, we report momentum-space measurements of electrons promoted to the graphene conduction band with visible light, and their subsequent relaxation.
  We observe a pronounced non-thermal distribution of nascent photoexcited electrons with lattice pseudospin polarization in remarkable agreement with results of simple tight-binding theory.
  By varying the excitation fluence, we vary the relative importance of electron-electron vs. electron-phonon scattering in the relaxation of the initial distribution.
  Increasing the excitation fluence results in increased noncollinear electron-electron scattering and reduced pseudospin polarization, although up-scattered electrons retain a degree of polarization.
  These detailed momentum-resolved electron dynamics in graphene demonstrate the capabilities of high-performance time-resolved momentum microscopy in the study of 2D materials and can inform the design of graphene devices.
\end{abstract}

The gapless, conical band structure of graphene is the origin of many exotic optical and electronic phenomena that can be harnessed for applications in optoelectronic devices. 
The broadband absorption \cite{nair2008} arising from the band linearity makes graphene suitable for optical sensors \cite{gao2021,xing2014} or photodetectors \cite{echtermeyer2014} in the near-IR to visible range and graphene is now routinely used in technologies such as saturable absorbers in passively mode-locked lasers.\cite{bao2009,bao2012,martinez2013,fu2014,cheng2020,goncalves2023,hasan2009}
The ultrafast optical response of graphene is due to carrier relaxation mediated by different scattering processes which lead to rapid thermalization of excited electrons.
Electron-electron (e-e) scattering is very efficient in graphene due to the gapless band structure and high carrier mobility, and also electron-phonon (e-ph) coupling between high-energy electrons and graphene's optical phonons is calculated to be exceptionally large\cite{malic2011}.
The microscopic details of excited carrier relaxation in graphene are thus both of fundamental interest and important for the design of graphene-based devices, and have been the subject of a substantial body of previous work.

Also proposed for use in graphene devices is an additional quantum number labeling Bloch wave functions in graphene termed lattice pseudospin \cite{min2008,san-jose2009}.
The lattice pseudospin $\phi$ refers to the relative phase of the wave function on the two equivalent carbon sublattices A and B in its honeycomb lattice via $\psi = \frac{1}{\sqrt{2}}\left[\psi_A \pm e^{i\phi}\psi_B\right]e^{i\mathbf{k}\cdot\mathbf{r}}$. 
Here the upper sign is for the conduction band, the lower sign is for the valence band, and $\mathbf{k}$ is the crystal momentum.
The angle $\phi$ also corresponds to the angular position of Bloch states around the $K$ points in the Brillouin zone,  with the pseudospin orientation parallel to the relative crystal momentum ($\mathbf{k} - K$) in the conduction band and antiparallel in the valence band \cite{cohen2016}.

 Photoexcitation of electrons from the valence band to the conduction band creates pseudospin polarization via the $\mathbf{k}$-dependence (and thus $\phi$-dependence) of the optical matrix elements \cite{malic2011}.
 Theoretical work has predicted that non-thermal electron distributions with strong lattice pseudospin polarization, i.e. the excited electrons of most interest for next-generation graphene-based optoelectronic devices, should be observable in pump/probe experiments in neutral graphene with low excitation fluence, with lifetimes of a few tens of femtoseconds.\cite{winzer2013,malic2012,malic2011,kadi2015}
Pseudospin dynamics in graphene have been previously studied using polarization-resolved optical spectroscopy \cite{mittendorff2014, yan2014,Yao_OptExp2015,trushin2015,konig-otto2016,danz2017, yan2017} and also using time- and angle-resolved photoemission (tr-ARPES).\cite{beyer2023,aeschlimann2017}
However, this previous work has been limited in scope.
The optical experiments have been performed with sufficient sensitivity to access the low-fluence regime, where e-ph scattering is expected to be the dominant relaxation mechanism. 
However these optical measurements have probed only limited energy ranges of the excited electron distribution, either at the initial excitation energy in degenerate pump/probe measurements or much higher or much lower energies in a few cases.
Furthermore, the optical observables involve drastic integrations over the momentum-space electron distributions, and require \emph{a priori} assumption of strict pseudospin selection rules for interpretation in terms of the electron dynamics around the Dirac cone.

Time-resolved ARPES experiments can measure the full excited electron distrubutions directly in momentum space and do not rely on any assumptions regarding selection rules, and there have been a number of previous tr-ARPES studies on graphene and graphite \cite{gierz2013,johannsen2013,johannsen2015,gierz2015,aeschlimann2017,na2019,duvel2022,keunecke2020,beyer2023}, with two studies addressing pseudospin polarization.\cite{aeschlimann2017,beyer2023} 
However, this previous tr-APRES work has been predominantly conducted at very high excitation fluence in the $\sim$mJ/cm$^2$ regime and/or on either heavily doped graphene or graphite, both of which have a large density of states (DoS) at the Fermi level $E_F$.
Both non-zero DoS at $E_F$ and strong pumping lead to the dominance of e-e scattering and very rapid thermalization of the electron distribution, and the data have been mostly well-described in terms of evolving Fermi-Dirac distributions characterized by time-dependent temperatures.\cite{aeschlimann2017, johannsen2013,johannsen2015, gierz2013, Na_PRB2020} 

In this Letter, we report tr-ARPES imaging of electrons excited with 2.4 eV ($\lambda$ = 517 nm) photons in high-quality neutral graphene samples produced via exfoliation.
A series of pump/probe measurements with high dynamic range and variable pump polarization are enabled by a unique high-performance instrument for time-resolved ARPES incorporating ultrashort extreme ultraviolet (XUV) pulses at 61 MHz repetition rate \cite{corder2018b, Corder_SPIE2018, kunin2023} with time-of-flight momentum microscopy\cite{Chernov_Ultramicroscopy2015,Medjanik_NatMat2017}.
High dynamic range enables us to observe pronounced non-thermal distributions with strong lattice pseudospin polarization.
For energies near the $h\nu_{\text{pump}}/2 = 1.2$ eV level populated by the pump pulse, at low excitation fluence the observed photoelectron signals are in remarkably good agreement with a simple model based on tight-binding theory accounting for the pump and probe optical matrix elements but with no consideration of e-e or e-ph scattering.
In the full distribution, pseudospin polarization is still visible down to 0.8 eV above the Dirac point, or multiple optical phonon energies below 1.2 eV, indicating the persistence of lattice pseudospin through multiple optical phonon scatterings.
With increasing excitation fluence we observe that increased e-e scattering leads to more rapid thermalization and reduced pseudospin polarization, although we do observe up-scattered electrons to retain a degree of polarization, as previously predicted by theory.\cite{malic2012}
To our knowledge, this is the first report of excited-state ARPES measurements in neutral graphene, and also the first to vary the excitation fluence over a range where e-e scattering vs. e-ph scattering are expected to be comparable\cite{winzer2013,jago2015}.

\begin{figure}[h]
  \includegraphics{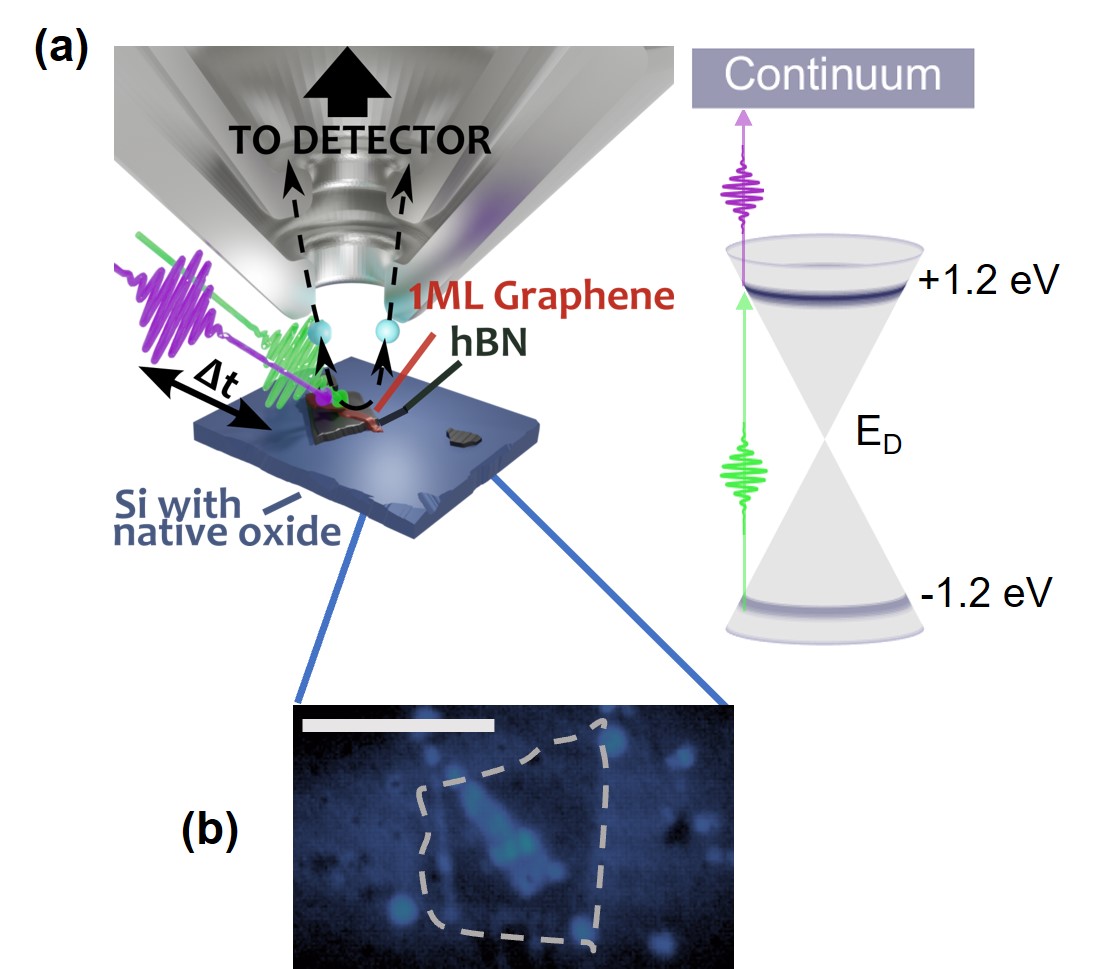}
  \caption{
    Experiment overview.
  (a) Linearly polarized pump pulses (green) promote electrons to 1.2 eV above the Dirac point (E$_\text{D}$) and a time-delayed XUV probe pulse ejects them into the continuum.
  (b) Real-space PEEM image of a graphene sample.
  The dashed lines show the region of the hBN support.
  The scale bar is 50 $\mu$m.
  }
  \label{fig:Intro} 
\end{figure}
Our overall experimental scheme is summarized in figure \ref{fig:Intro}.
Visible ($h\nu_{\text{pump}} = 2.4$ eV) pump pulses with variable polarization and $p$-polarized XUV (20-30 eV) probe pulses impinge on the sample at 48 degrees.
Photoelectrons are collected with a custom time-of-flight momentum microscope similar to that described by Medjanik et al.\cite{Medjanik_NatMat2017}
Samples used in this work were prepared by exfoliation, and consist of monolayer graphene stacked on a buffer layer of hexagonal boron nitride (hBN) stacked on a silicon substrate with its native oxide layer.
A representative real-space photoelectron microscopy (PEEM) image one such sample is shown in figure \ref{fig:Intro}b.
The procedure for fabricating samples and additional characterization are presented in the supporting information.
We repeated our experiments on several graphene samples and over a range of XUV photon energies, with the key results reported here reproducing across several samples and a range of conditions.
The main difference observed between different samples is the background level (noise floor) observed in the photoelectron spectra, with higher background precluding clear observation of non-thermal distributions in experiments with lower excitation fluence. Our laser system, cavity-enhanced high harmonic generation (CE-HHG) source, and beamline have been described in detail previously.\cite{li2016a,corder2018b,Corder_SPIE2018}.
We select photoelectrons emerging from the graphene sample using a small aperture in a real-space image plane of the momentum microscope, and also high-pass filter the energy distribution (EDC) as described previously by Kunin et al.\cite{kunin2023}.
The XUV probe beam is 24 $\times$ 16 $\mu$m$^2$ FWHM on the sample, and the pump beam size is set to be at least three times as large such that the recorded sample area is uniformly pumped.
All measurements are performed at a base pressure of $\sim$$5 \times 10^{-10}$ Torr and with the sample held at room temperature. 
\begin{figure*}[t]
  \includegraphics{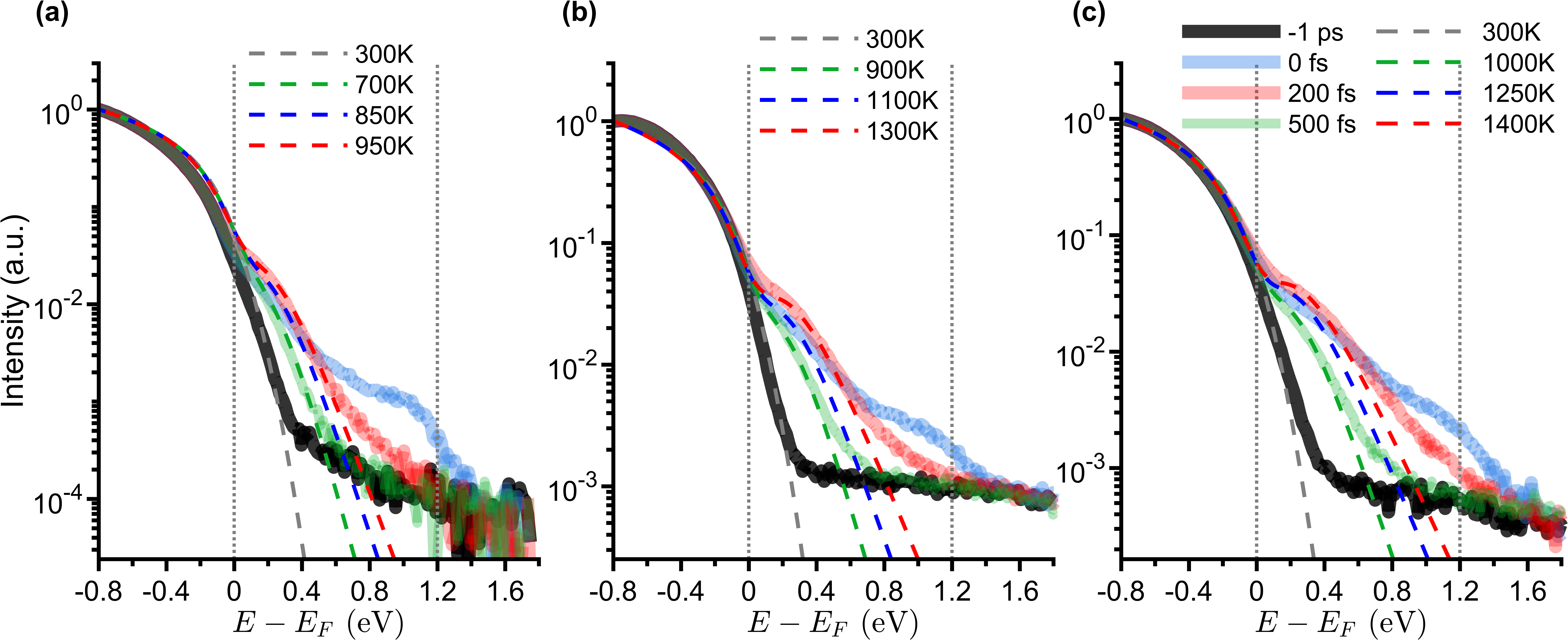}
  \caption{
  Energy distribution curves (EDC, solid lines) at selected time delays for (a) 45 $\mu$J/cm$^2$, (b) 132 $\mu$J/cm$^2$, and (c) 207 $\mu$J/cm$^2$. 
  Simulated thermal distributions at different temperatures (dashed) are shown for comparison.
  The experimental data are at time delays $\Delta t = -1$ ps in black, $\Delta$t = 0 fs in blue, $\Delta t = 200$ fs in red, and $\Delta t = 500$ fs in green. 
  The vertical dotted lines in gray represent the Dirac point $E - E_F = 0$ and $E-E_F =h\nu_{\text{pump}}/2 = 1.2$ eV for the direct excitation energy.
  The noise floor of the measurements at high energies is seen in the negative delay data (black curves).
  }
  \label{fig:electemp}
\end{figure*}

We determine the time-zero (delay where pump and probe pulses maximally overlap on the sample) and instrument response function (IRF = cross-correlation between pump and probe) using both the graphene signal itself, and also complementary pump/probe experiments on an Au(111) crystal. %, as described in detail in the supporting information.
The IRF is well-described as a Gaussian with a 200 fs FWHM.
Here we report measurements with incident pump fluences between 45 and 207 $\mu$J/cm$^2$.
We correct for small time-dependent surface photovoltage shifts (due to excitation of the silicon substrate) less than 40 meV via observing shifts in the Dirac point, and overall determine the position of the Dirac point with an uncertainty of 50 meV.
From comparing ground state signals to those expected from a Fermi-Dirac distribution at 300 K, we estimate the energy resolution of these experiments to be approximately \Eres \; FWHM.

Figure \ref{fig:electemp} shows momentum-integrated EDCs on a logarithmic scale for three different fluences: 45, 132, and 207 $\mu$J/cm$^2$.
For comparison, thermal signals expected from Fermi-Dirac distributions at different temperatures are shown by the dashed lines.
The thermal signals are constructed via the product of the Fermi function $\braket{n(E)} = (e^{(E - E_F)/k_B T} + 1)^{-1}$ with the density of states and photoemission matrix elements derived from tight-binding theory, all convolved with a Gaussian of \Eres \; FWHM to account for the instrumental energy resolution. 
The kinks in the thermal signals at $E = E_F$  are due to the zero in the density of states at the Dirac point.
At delays ($\Delta t$) longer than our IRF width, i.e. delays where the pump and probe pulses no longer overlap significantly, the spectra are well described with the simulated thermal signals with temperatures consistent with previous work.\cite{Na_PRB2020,johannsen2013}
Note that it is expected that the temperature does not scale linearly with the fluence due to both the non-constant graphene DoS and the onset of saturated absorption in this fluence regime.\cite{goncalves2023}
For all fluences, the momentum- and energy-integrated signal for all electrons above the Dirac point is fit well by a biexponential decay with $\tau_1 \approx 200$ fs and $\tau_2 = 1-3$ ps depending on fluence, as shown in figure S3 of the supporting information.

However, near $\Delta t =0$, where the pump and probe pulses overlap, significant population is observed above 0.8 eV that cannot be described by a Fermi-Dirac distribution.
The behavior we observe in the EDC agrees qualitatively with that predicted by Winzer et al.\cite{winzer2013}, who performed density-matrix/Bloch equation simulations of electron dynamics in graphene at different excitation fluences.
%Jin sentence: At low fluence, predict that a phonondominated scattering process occurs in steps of constant phonon energies in very low excitation regime, resulting in the discrete steps in the electron distribution [20].
At low fluence, electrons relax by emitting optical phonons in discrete steps and the EDC shows a plateau behavior with a steep drop-off at $h\nu_{\text{pump}}/2$, with few electrons above this initial excitation energy.
At higher fluence, electron-electron scattering is dominant and the distribution is smoother with a tail extending to higher energies well above $h\nu_{\text{pump}}/2$.
\begin{figure}[h!]
  \includegraphics{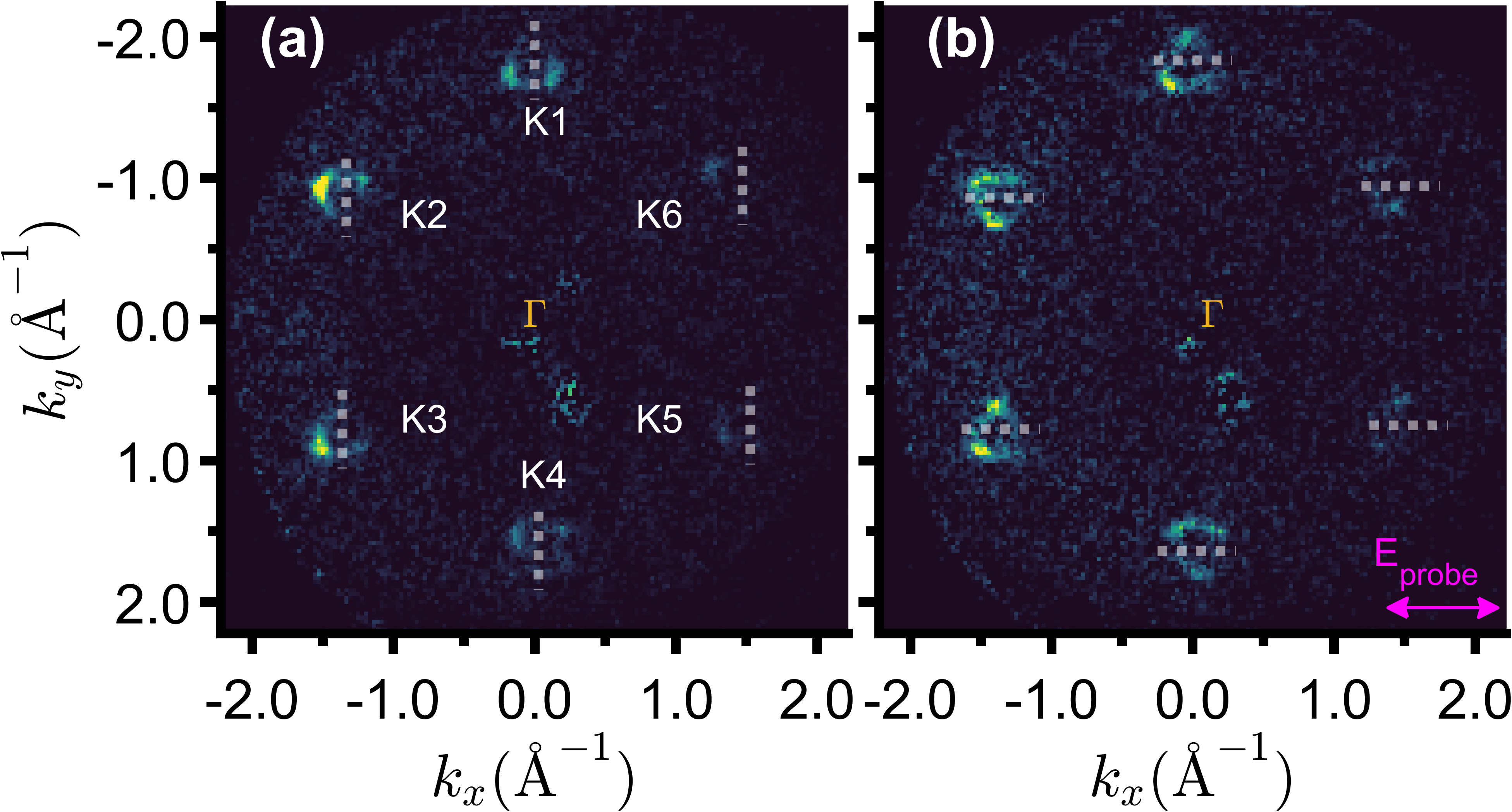}%Fig3_InitialDist_NanoLett_Remove3D.png}
  \caption{\textbf{Polarization-dependent initial pseudospin anisotropy.} 
  Momentum distributions for electrons between 1.05 and 1.21 eV above the Dirac point at $\Delta t = 0$ are shown for (a) $y$- and (b) $x$-polarized pump excitation.
  The excitation fluence is 45 $\mu$J/cm$^2$,
  and the directions of pump and probe polarization are indicated by the dashed white lines and magenta arrow, respectively. 
  The intensity near the $\Gamma$ point is an artifact of the detector.}
  \label{fig:kIm_p_s}
\end{figure}
\begin{figure*}[t]
  \includegraphics{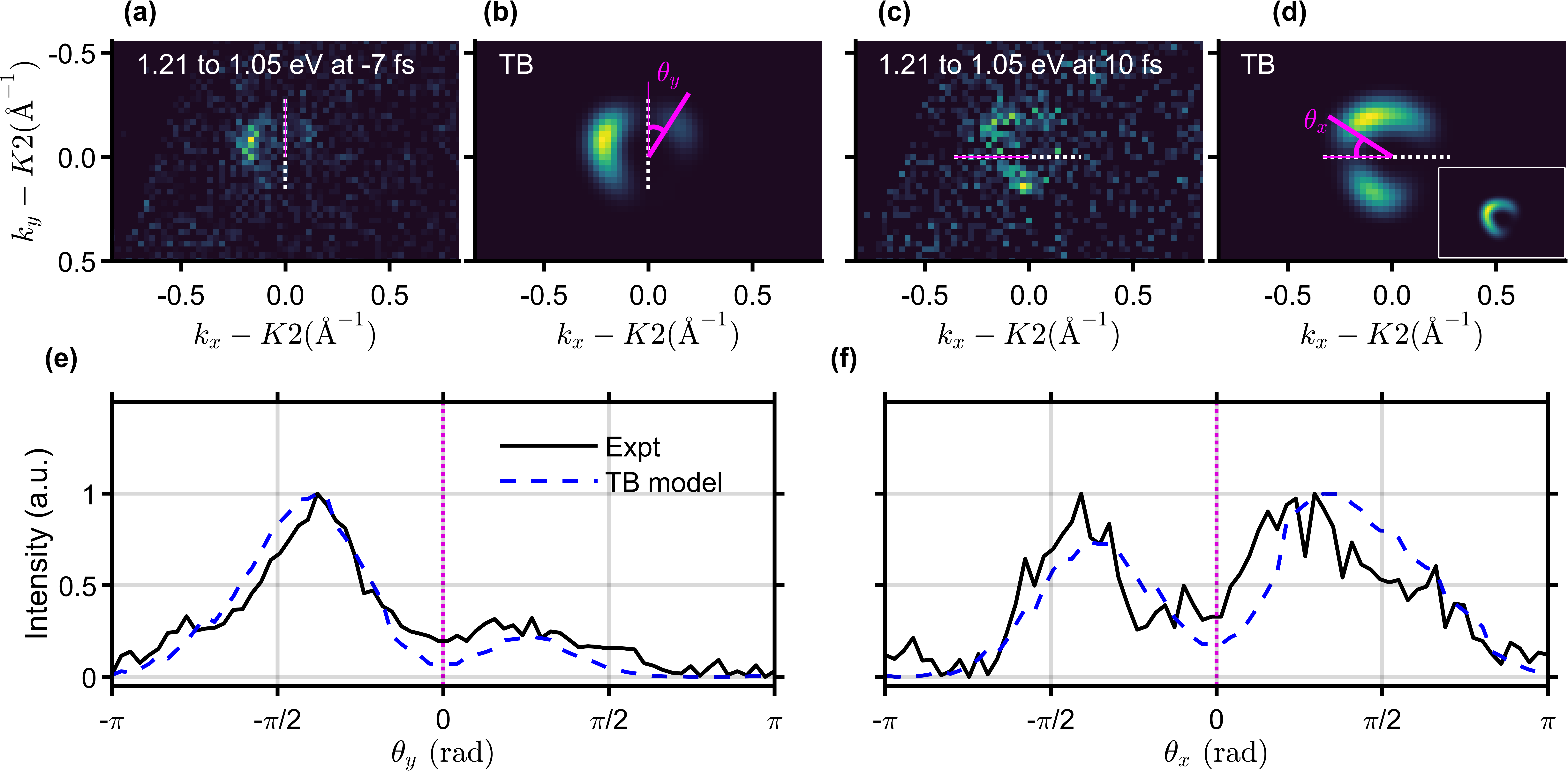}
  \caption{
  Comparison to tight-binding (TB) theory.
  (a) and (c) K2 Momentum distribution spectra for electrons from 1.05 to 1.21 eV above the Dirac point for $\Delta t = 0$ for  $y$- and $x$-polarized excitation, respectively. 
  (b) and (d) Photoemission momentum maps predicted by the TB model according to equation (\ref{eqn:TBmodel}).
  Inset in (d) is equation (\ref{eqn:TBmodel}) without the pump matrix element, illustrating the so-called ``dark corridor'' and how the signals would look in the absence of pseudospin polarization. 
  Comparison of angle-resolved intensities between the TB model and the experiment for (e) $y$-polarization and (f) $x$-polarization. 
  The pump polarization direction is indicated by the white dotted lines in the top panels. 
  The assigned pump nodes are denoted by the thin magenta lines in the top panels and by magenta dotted lines in the bottom panels. 
  \label{fig:compareTB}
}
\end{figure*}

Figure \ref{fig:kIm_p_s} shows the photoelectron momentum distributions for electrons between 1.05 and 1.21 eV above the Dirac point, i.e. within one optical phonon of the initial excitation at $h\nu_{\text{pump}}/2$, with the in-plane component of the excitation electric field polarized in the $x$ and $y$ directions.
For this data the excitation fluence is 45 $\mu$J/cm$^2$ and the pump and probe pulses are maximally overlapped ($\Delta t = 0$).
Nodes, illustrated by the white dashed lines, are clearly observed along the pump polarization direction, as expected from the optical matrix element pseudospin selection rules.\cite{malic2011} 
Also visible are the so-called dark corridors along the $\Gamma-K$ direction and a left-right asymmetry in the images due to the $\mathbf{k}$-dependence of the photoemission matrix element and the $p$-polarization of the XUV light.\cite{gierz2011,shirley1995}
In what follows, we do not include data from the $K5$ and $K6$ regions in our analysis due to the low statistics in these regions.

In figure \ref{fig:compareTB} we compare the observed photoelectron distribution recorded at 45 $\mu$J/cm$^2$ excitation fluence to a simple model
\begin{equation}\label{eqn:TBmodel}
\begin{aligned}
  I(\mathbf{k}&) \propto \int_{1.05 \hspace{0.05em} \text{ eV}}^{1.2 \hspace{0.2em} \text{eV}} d\varepsilon
  \big[ \; \delta(\varepsilon - \mathcal{E}_{\text{TB}}(\mathbf{k})) \times \\
  & |\mathbf{M}_{\text{pump}}(\mathbf{k})|^2 \times
  |\mathbf{M}_{\text{probe}}(\mathbf{k},\mathbf{k}_z)|^2
  \; \big ]
  \ast h(\mathbf{k})\;,
  \end{aligned}
  \end{equation}
  %
  \iffalse
\begin{align}\label{eqn:TBmodel}
  I(\mathbf{k},E) \propto \int_{1.05 \text{ eV}}^{1.2 \text{eV}} d\varepsilon
  \left[\;
  \delta(\varepsilon - \mathcal{E}_{\text{TB}}(\mathbf{k})) \times 
  |\mathbf{M}_{\text{pump}}(\mathbf{k})|^2 \times
  |\mathbf{M}_{\text{probe}}(\mathbf{k},\mathbf{k}_z)|^2 
  \; \right]
  \ast h(\mathbf{k})\;,
\end{align}
\fi
%
where $\mathcal{E}_{\text{TB}}(\mathbf{k})$ is the band dispersion, $\mathbf{M}_{\text{pump}}(\mathbf{k})$ is the dipole matrix element between the valence band and conduction band, and $\mathbf{M}_{\text{probe}}(\mathbf{k},\mathbf{k}_z)$ is the dipole matrix element between the conduction band states and plane wave final states with wave vector $\mathbf{k}_{\text{tot}} = \mathbf{k} + \mathbf{k}_z$, all derived from tight-binding theory.\cite{malic2011,shirley1995}
$\hbar^2 |\mathbf{k}_{\text{tot}}|^2/2m = h\nu_{\text{XUV}} - W - \varepsilon$, with $W \approx 5$ eV the work function.
The function $h(\mathbf{k})$ used for convolution is a 2D Gaussian of 0.11 \AA$^{-1}$ FWHM reflecting the momentum resolution of our measurements, determined from the size of our observed Dirac point in momentum space images (this also effectively captures the energy resolution).
The simulated distributions for $y$-polarized and $x$-polarized light are shown in figures \ref{fig:compareTB}b and \ref{fig:compareTB}d, respectively for the region labeled $K2$ in figure \ref{fig:kIm_p_s}, alongside zoomed-in images of the momentum-space photoelectron distributions from the experiment.
In figures \ref{fig:compareTB}e and \ref{fig:compareTB}f we compare theory and experiment quantitatively by plotting the intensity vs. angle around $K$. 
Remarkable agreement is seen with only a single overall scaling parameter applied to the theoretical curves to match with the experiment, indicating that these electrons have undergone minimal scattering processes that alter their pseudospin polarization.
Similar results are observed for $K3$.
Somewhat worse agreement with the simple tight-binding model is observed for $K1$ and $K4$, likely due to reduced accuracy of modeling the photoemission matrix element with tight binding theory for photoelectron final-state momenta further away from the light polarization vector, as we also observe larger discrepancies between ARPES momentum maps and tight-binding theory for the occupied ground-state valence band for $K1$ and $K4$. Theory/experiment comparisons for $K1$, $K3$, and $K4$ are presented in the supporting information.

\begin{figure}[H]
  \includegraphics{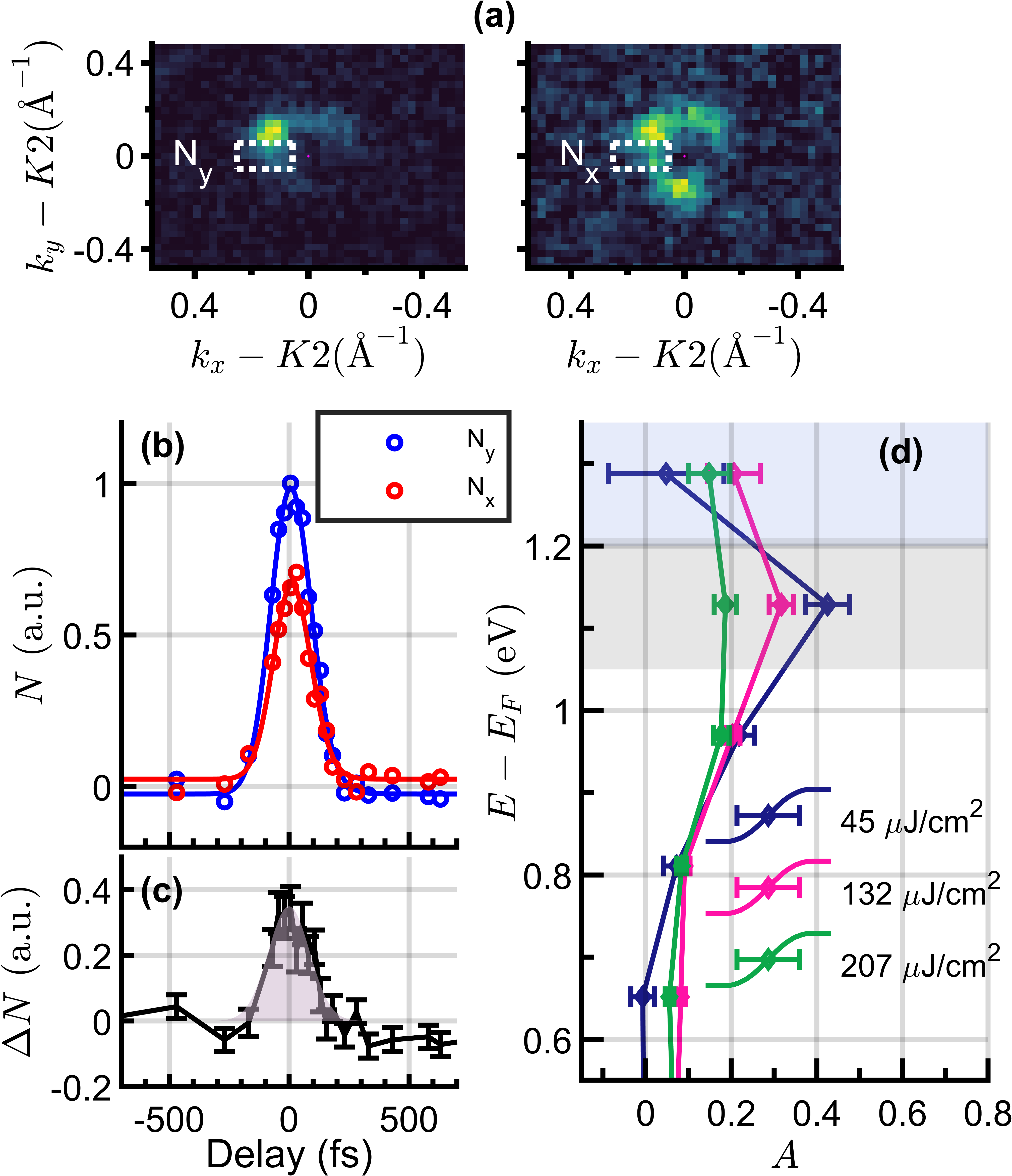}%IvsPhi_CompareGd_TB_K2_mod.jpg}%IvsPhi_K2.png} %IvsPhi_CompareGd_TB_K2_mod.jpg
  \caption{
  Pseudospin Relaxation.
  (a) ROI illustration for K2. Data are recorded with $y$-polarized (left) and $x$-polarized (right) pump pulse. 
  Data from similar ROIs in all the valleys are combined to produce the transient signals $N_y$ and $N_x$ integrated over all energies $>$ 0.8 eV.   
  (b) $N_y$ (blue) and $N_x$ (red) vs. pump/probe delay.
  (c) The difference $\Delta N = N_y - N_x$ (black) and the IRF (shaded gray).
  Both the populations and the anisotropy track the IRF, indicating relaxation much faster than the 200-fs IRF.  
  (d) Normalized anisotropy $A$ vs. energy for different fluences. At higher fluences anisotropy is overall reduced but persists through more scattering events.
  More details in text. 
  \label{fig:pseudorelax}
  }
\end{figure}

To investigate the dynamics of the momentum anisotropy we performed measurements with pump polarization alternating between $x$ and $y$ directions.
Figure \ref{fig:pseudorelax}b shows the photoelectron signals vs. pump/probe delay, recorded in selected regions of interest (ROI) oriented 90 degrees to the nodes created by $y$-polarized pump pulses.
An example ROI for $K2$ is shown in figure \ref{fig:pseudorelax}a. 
ROI transient signals with $x$-polarized and $y$-polarized pump pulses are labelled $N_x$ and $N_y$ respectively.
Figure \ref{fig:pseudorelax}c shows their difference $\Delta N = N_y - N_x$.   
Using the same ROI while alternating the excitation polarization between $x$ and $y$ each pump/probe scan ensures that both datasets share the same probe matrix element and also any systematics due to detector response inhomogeneity, enabling careful comparison of the intensities within the ROI.
The ROI signals are integrated over all energies of the non-thermal distribution, i.e. 0.8 eV and above, and the shaded gray Gaussian is the temporal IRF.
As seen in figures \ref{fig:pseudorelax}b and \ref{fig:pseudorelax}c, both the excited population ($N_x$ or $N_y$) and the anisotropy $\Delta N$ closely follows the 200 fs instrument response, indicating very fast relaxation.

Although we cannot recover anisotropy relaxation times from the pump/probe traces, significant information about the dynamics comes from analyzing the energy dependence of the anisotropy.
Since electrons are initially promoted to 1.2 eV above the Dirac point by the pump pulse, electrons observed away from this energy get there via either e-e or e-ph scattering. 
Figure \ref{fig:pseudorelax}d shows the energy dependence of the normalized anisotropy $A \equiv (N_y-N_x)/(N_x + N_y)$ at $\Delta t = 0$ for different excitation fluences.
The maximum anisotropy is always observed at the initial excitation energy corresponding to the gray energy bin of figure \ref{fig:pseudorelax}d, and reduced anistoropy is observed at all other energies populated by scattering.
The energy binning of 160 meV is chosen to correspond to the lowest optical phonon energy in graphene\cite{Mohr_PRB2007,na2019} such that the points in figure \ref{fig:pseudorelax} are spaced by approximately one optical phonon energy.
As predicted by theory, at low fluence where e-ph coupling is the main relaxation mechanism, relaxation of the pseudospin polarization via e-ph interactions is very efficient. \cite{malic2012}
The low fluence data in figure \ref{fig:pseudorelax}d show that the observed momentum anisotropy of the electrons approximately halves for each optical phonon emitted.

In contrast, e-e scattering is expected to better preserve the momentum anisotropy due to pseudospin dependence of the e-e interaction $V \propto 1 + e^{i(\phi_f -\phi_i)}$, which favors collinear scattering $\phi_f = \phi_i$.\cite{malic2011,malic2012}
At higher fluences, where e-e scattering dominates, we do observe that the anisotropy of relaxed electrons is comparable to what we observe at 1.2 eV, however we also observe that the anisotropy is overall reduced with increasing fluence as shown in figure \ref{fig:pseudorelax}c, indicating significant non-collinear scattering.\cite{aeschlimann2017,winzer2013,konig-otto2016}  
The same degree of anisotropy is observed for electrons at energies above 1.2 eV, which appear with significant population for our two higher fluences.
These electrons are expected to be upscattered only via e-e scattering since the initial optical phonon population at room temperature is small.\cite{winzer2013}

In conclusion, in this Letter we have reported momentum-space observations of optically excited electrons in graphene with large lattice pseudospin polarization and a pronounced non-thermal distribution.
Such distributions resulting from optical excitation have been predicted by theory, but not observed in previous graphene tr-ARPES experiments due to experimental conditions and instrument performance.
Conducting the experiment with variable excitation fluence and in neutral graphene, with zero DoS at the Fermi level, enables us to control the relative importance of e-e and e-ph scattering in electron relaxation.
We find that under low fluence excitation, photoelectrons observed at the excitation energy have a distribution in remarkable agreement with tight-binding theory completely neglecting e-e and e-ph scattering.
At higher excitation fluences, e-e scattering rapidly redistributes the population, but still preserves the anisotropy to a degree.
Further detailed modeling of the fluence-dependent anisotropy vs. energy data may extract mode-specific electron relaxation rates but this is beyond the scope of this Letter.
This work demonstrates the abilities of high-performance momentum microscopy for excited-state imaging and can inform the design of graphene-based optoelectronic devices. 
\begin{suppinfo}
The supporting information contains details about sample preparation, 
full datasets of EDC vs. pump/probe delay and related fitting of overall decay of the integrated electron signals,
procedures for determining the IRF,
and comparisons between the experimental data and tight-binding theory for valleys $K1$, $K3$ and $K4$. 

\end{suppinfo}  

\begin{acknowledgement}
  This material is based upon work supported by the U.S. Department of Energy, Office of Science, Office of Basic Energy Sciences under award number DE-SC0022004 and the Air Force Office of Scientific Research under FA9550-20-1-0259.
  R.K.K. acknowledges support from the U.S. National Science Foundation under Grant No. CHE-1935885.
  X.D. acknowledges support from the U.S. National Science Foundation under Grant No. DMR-1808491.
  M.G.W. acknowledges support from the U.S. Department of Energy (DOE), Office of Science, Office of Basic Energy Sciences, Chemical Sciences, Geosciences, and Biosciences (CSGB) Division, and the Catalysis Science Program under DOE Contract No. DE-SC0012704. 
  Z.H.W. acknowledges support from the U.S. National Science Foundation Graduate Research Fellowship Program.

\end{acknowledgement}

\bibliography{Gr_main}

\end{document}